\documentclass[authoryear,preprint,review,12pt]{elsarticle}

\usepackage{amssymb}
\usepackage{graphicx}
\usepackage{caption}
\usepackage{amsmath}

\def\astrobj#1{#1}
\journal{New Astronomy}

\begin{document}

\begin{frontmatter}



\title{The low mass ratio contact binary system \astrobj{V728~Herculis}}


\author[abc,def]{N. Erkan\corref{dip}}
\author[xyz]{B. Ula\c{s}}
\address[abc]{Department of Physics, Faculty of Arts and Sciences, \c{C}anakkale Onsekiz Mart University, Terzio\~{g}lu Kamp\"{u}s\"{u}, TR-17020, \c{C}anakkale, Turkey}
\address[def]{Astrophysics Research Centre and Observatory, \c{C}anakkale Onsekiz Mart University, Terzio\~{g}lu Kamp\"{u}s\"{u}, TR-17020, \c{C}anakkale, Turkey}
\address[xyz]{\.{I}zmir Turk College Planetarium, 8019/21 sok. No:22, \.{I}zmir, Turkey}
\cortext[dip]{Corresponding author \\
E-mail address: nacierkan@comu.edu.tr}

\begin{abstract}
We present the orbital period study and the photometric analys of the contact binary system \astrobj{V728~Her}. Our orbital period analysis shows that the period of the system increases ($dP/dt=1.92\times10^{-7}$dyr$^{-1}$) and the mass transfer rate from the less massive component to more massive one is $2.51\times10^{-8}$ M$_{\odot}y^{-1}$. In addition, an advanced sinusoidal variation in period can be attributed to the light--time effect by a tertiary component or the Applegate mechanism triggered by the secondary component. The simultaneous multicolor $BVR$ light and radial velocity curves solution indicates that the physical parameters of the system are $M_1=1.8M_{\odot}$, $M_2=0.28M_{\odot}$, $R_1=1.87R_{\odot}$, $R_2=0.82R_{\odot}$, $L_1=5.9L_{\odot}$, and $L_2=1.2L_{\odot}$. We discuss the evolutionary status and conclude that \astrobj{V728~Her} is a deep ($f$=81\%), low mass ratio ($q$=0.16) contact binary system.
\end{abstract}


\end{frontmatter}

\section{Introduction}

\astrobj{V728~Her} is a member of the group of low mass ratio ($q\leqslant$0.25) contact binary systems. The light variation of the system was first detected by \cite{kur77}. The author calculated the orbital period of the system as 0$^d$.44625. \cite{nel88} proposed that the spectral type of the system is F3. The authors observed the light curves of the system in three filters. They also improved the orbital period value to 0$^d$.471302. The $BV$ light curves were published by \cite{age88} who redetermined the equatorial coordinates of the system. \cite{sam89} presented photoelectric $B$ and $V $ light curves. A light curve solution and a period study of \astrobj{V728~Her} was introduced by \cite{sam90}. The author noticed the mass transfer from the secondary to primary component. \cite{nel95} analyzed the $BVI_{c}$ light curves and  radial velocity curve of the system by assuming two different models. They concluded that the system is a contact binary whose components have convective envelopes. \cite{nel99} investigated the change in the orbital period and emphasized the probability of sudden or gradual increase. The system was listed by \cite{pri03} and \cite{get06} in their field contact binary catalog and bright contact binary catalog, respectively. \cite{bra05} noticed the asymmetric behavior of the light curve which could not be seen in many previous studies. \cite{chr12} noted that the previous analyses are not very reliable because of the individual peculiarities in the light of the system. Finally, \cite{yan15} included the system to their statistical study of 46 deep, low mass ratio overcontact systems.

In the next section, details of our observations are explained. An investigation of variation in the orbital period of the system is presented in Sec. 3. In Sec. 4, the simultaneous analyses of light and radial velocity curves are presented. We discuss the results and give the concluding remarks in the last section.

\section{New Observations}

\begin{figure}
\centering
\includegraphics[scale=1, angle=0]{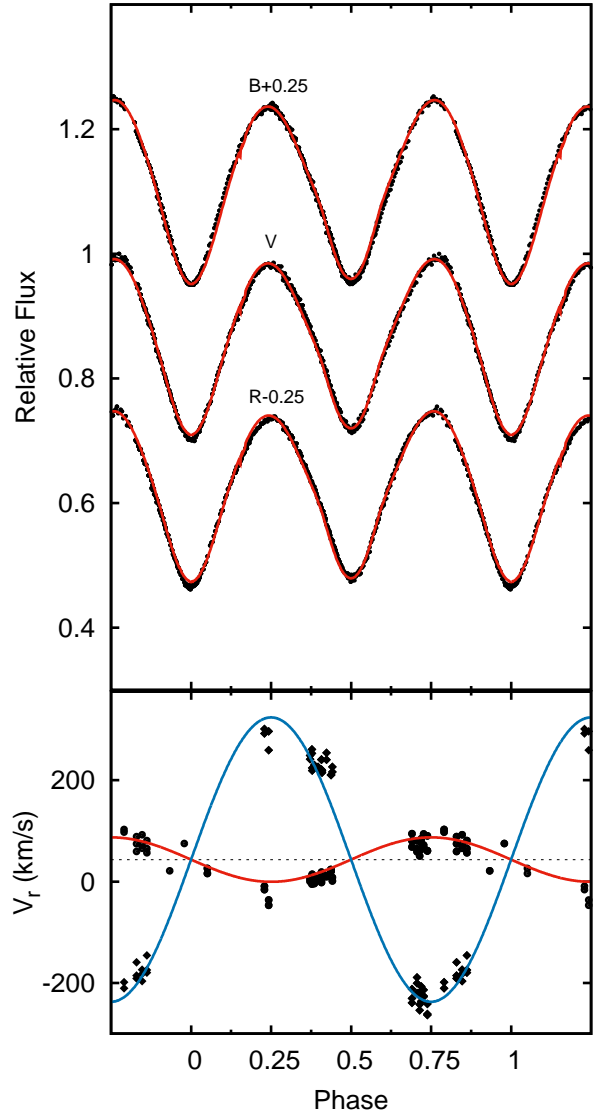}
\caption{Theoretical and observational light (top) and radial velocity (bottom) curves of V728 Her. Theoretical curves are shown by solid lines. Radial velocity data are taken from \cite{nel95}.}
\label{figlc}
\end{figure}

CCD observations of the system were made in Cousins/Bessel $B$, $V$ and $R$ filters attached to the 1.22-m telescope of \c{C}anakkale Onsekiz Mart University Observatory. The observational data cover 5 nights between June and July 2013 (HJD~2456451.4422 to HJD~2456489.5570). 409  points in $B$ filter, 378 in $V$ filter and 401 in $R$ filter were collected during the observations. Our mean photometric errors are 0.003 in $B$ and $V$, 0.002 in $R$ filters. The comparison and check stars were chosen as TYC~3081-1028-1 and TYC~3081-571-1 respectively. Observational light curves of the system are plotted in Fig.~\ref{figlc}. Table~\ref{tabmin} lists two times of minimum light derived by using Kwee--van Woerden method \citep{kwe56}.

The light curve of the system shows magnitude difference between two maxima. The primary and the secondary minima are round bottomed and their depths are very close to each other, -- 0$^m$.38 and -- 0$^m$.37, respectively. A slight asymmetry, which is previously mentioned by \cite{sam90}, can be seen in the phase interval  between 0.15 and 0.25. 

\section{Orbital Period Analysis}
\cite{age88} improved the orbital period value of the system by using the least--squares method. \cite{nel95} obtained eight minimum times and applied an orbital period analysis. Eighty times of minimum light were collected and analyzed by \cite{sam90}. The author indicated a drastic period increase in the $O-C$ curve. \cite{nel99} fitted the $O-C$ curve using the least--squares method and represented the curve by both straight line and parabola. The author noted the possibility of sudden period increase at about 2479th cycle. \cite{nel99} also mentioned that the variation could be attributed to a gradual increase in the orbital period. The times of minimum light which are obtained after the year 2000 cleared the behavior of the $O-C$ curve.

\begin{figure}
\centering
\includegraphics[scale=0.7, angle=90]{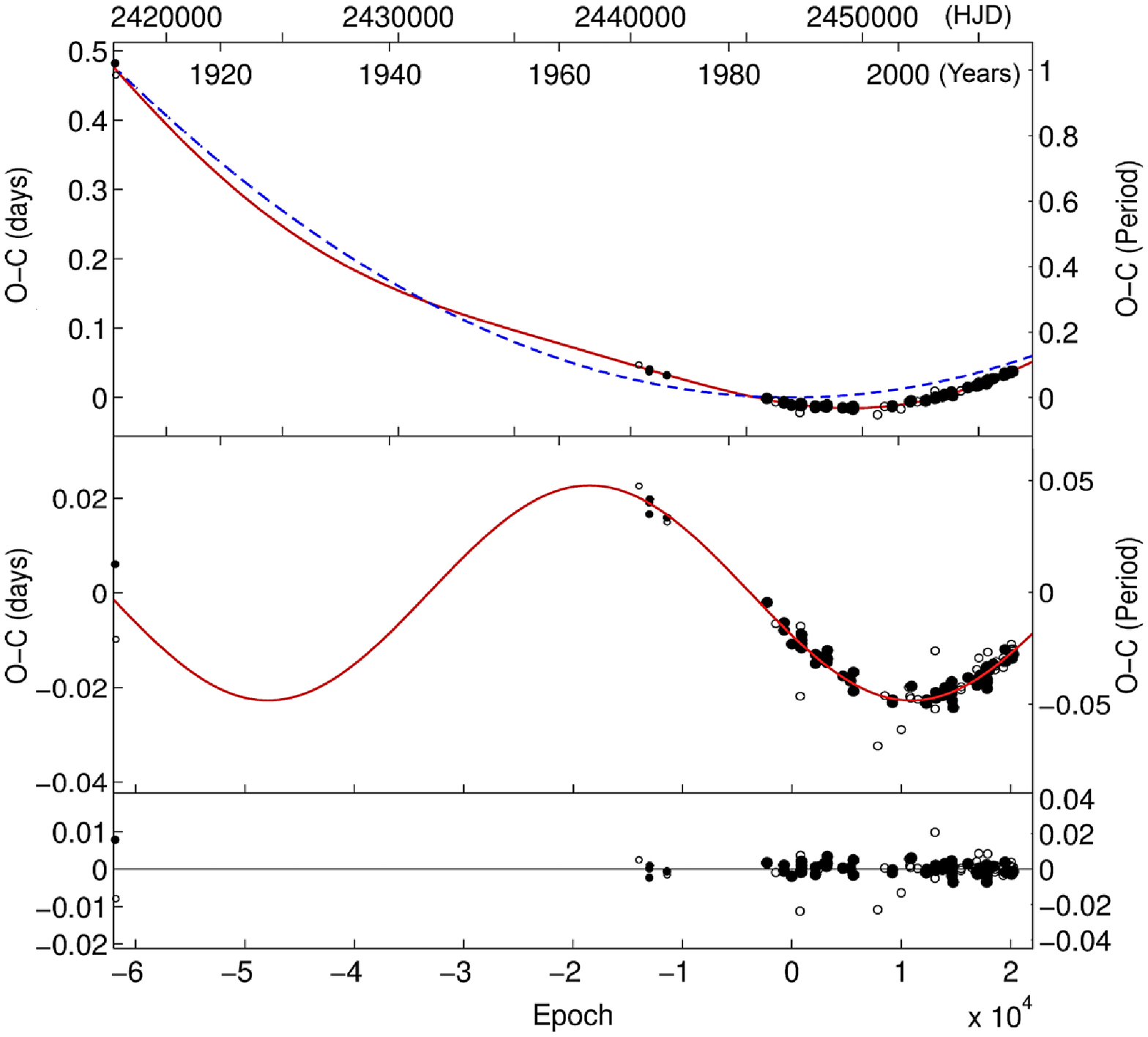}
\caption{Plot of the period analysis. The dashed and solid lines in the uppermost panel show the parabolic variation and the sinusoidal variation superposed on the parabola. The middle panel shows the sinusoidal fit after removal of the upward parabolic variation. We represent in the lowest panel the final residuals yielded after the analysis.}
\label{figoc}
\end{figure}

We obtained two minimum times (Table~\ref{tabmin}) and collected 121 times of minimum light from the database of Czech Astronomical Society$\footnote{http://var.astro.cz}$.  Therefore, the orbital period change of the binary was investigated by using 123 times of minimum light in total. Since the main shape of the $O-C$ curve (Fig.~\ref{figoc}) is an upward--parabola, mass transfer from the less massive (secondary) component to the more massive one is expected. In addition, a sinusoidal variation superposed on the parabola can also be seen in the $O-C$ curve. The sine--like variation in which both the primary and the secondary minima follow the same trend can mainly occur as a result of two different physical phenomena: (i) the light--time effect which is observed because of the presence of an external third body and (ii) the Applegate mechanism which is generally seen in magnetically active components of binary stars. 

We first analyze the data by combining mass transfer and third body assumptions. The {\tt LITE} code \citep{zas09} which is based on the simplex algorithm was used to calculate the resulting parameters. The code solves the input data to represent it by following formula :

\begin{eqnarray*}
 \text{HJD(MinI)} &=& T_{0} + P_{0} \times \text{E} + Q \times \text{E}^{2} \\
                   && + {\frac{a_{12}\sin i^{\prime}} {c}} [{ \frac{1-{e^{\prime}}^2} {1+e^{\prime}\cos \upsilon^{\prime}} } \sin (\upsilon^{\prime}+\omega^{\prime}) \\ 
                   && + e^{\prime}\sin \omega^{\prime}]
\end{eqnarray*}
where $T_{0}$ is the epoch for the primary eclipse, $P_{0}$ is the orbital period and  $E$ is the integer eclipse cycle number of the binary system. The orbital parameters of the tertiary component are the semi--major axis $a_{12}$, the inclination of the eclipsing pair about the third body $i^{\prime}$. $\upsilon^{\prime}$ refers to true anomaly of the position of the center of mass. The sinusoidal variation shows that the orbital eccentricity of the third body ($e^{\prime}$) is zero, therefore, the longitude of the periastron of the binary ($\omega^{\prime}$) and the time for periastron passage of the tertiary component ($T{^{\prime}}_0$) are undefined.

The $O-C$ curve and the final fit of the solution are shown in Fig.~\ref{figoc}. The result of the analysis shows that the increment in the orbital period is $dP/dt=1.92\times10^{-7}$ dyr$^{-1}$. The increase in the period implies a mass transfer rate of $dm / dt=2.51\times10^{-8}$ M$_{\odot}$yr$^{-1}$ according the formula given by \cite{sin86} who assumed that the mass transfer between the components is conservative. Additionally, a third companion with an orbital period of 76 years can be assigned to the sinusoidal variation. The results of the solution are listed in Table~\ref{tabth}. The probable mass values of the tertiary component were computed as 0.9, 0.5 and 0.4 M$_{\odot}$ for inclination values of 30$^{\circ}$, 60$^{\circ}$ and 90$^{\circ}$, respectively.

Addressing the sine--like variation to the Applegate mechanism, on the other hand, requires to calculate the subsurface magnetic fields of the components. \cite{app92} suggested that some modulations in the orbital periods of eclipsing binaries are observed because of the interactions between orbit and the shape of the magnetically active component. The Applegate mechanism requires that the variation in luminosity and differential rotation must confirm the criteria levels of $\Delta L / L \approx 0.1$ and $\Delta \omega / \omega \approx 0.01$.  We calculated that these parameters are $\Delta L_{1} / L_{1} = 0.001$, $\Delta \omega_{1} / \omega_{1} = 0.0002$ and $\Delta L_{2} / L_{2} = 0.021$, $\Delta \omega_{2} / \omega_{2} = 0.0024$ for the primary and the secondary component according to the formulae given by \cite{app92}. Furthermore, the required subsurface magnetic field of the secondary component was found to be larger ($B_1$ = 2.78 kG, $B_2$ = 5.35 kG), therefore, in this model the less massive star undergoes the Applegate mechanism and is responsible for the cyclic period variation.

\section{Simultaneous Solution of Light and Radial Velocity Curves}

The first light curve analysis of the system was made by \cite{sam90} who applied Wilson--Devinney program \citep{wil71} on the normal points of $BV$ light curves. The author pointed out that the system is W UMa type with two components having almost equal temperatures ($\Delta$T$\approx$100~K). \cite{sam90} also mentioned that the fillout factor of the system is close to 20$\%$. \cite{nel95} analyzed their phase--binned $BVI_{c}$ light and radial velocity curves with radiative and convective atmosphere model approximations separately. They concluded that the convective model improved the results considerably. According to their results, the temperature difference between components is about 150~K.

The $BVR$ light curves of the system were combined with the radial velocity data given by \cite{nel95} in our study. A simultaneous solution was applied to input data. The {\tt PHOEBE} \citep{prs05} software, which is based on the Wilson--Devinney code \citep{wil71}, was used during the analysis. Albedo values A$_1$,A$_2$ and gravity darkening coefficients g$_1$, g$_2$ were adopted from \cite{ruc69} and \cite{ham93}, respectively. The temperature of the primary component was set to 6000~K which is very close to the value of \cite{nel95}. The adjustable parameters of the solution were the inclination $i$, mass ratio $q$, velocity of the center of mass $V_0$, semi--major axis $a$, surface potential $\Omega_1 = \Omega_2$, temperature of secondary component $T_2$, luminosity of the primary component $L_1$, time of minimum light $T_0$ and the orbital period $P$. Since the light curve shows W~UMa type characteristics the appropriate running mode for contact binaries was chosen. 

\begin{figure}
\centering
\includegraphics[scale=1.5, angle=0]{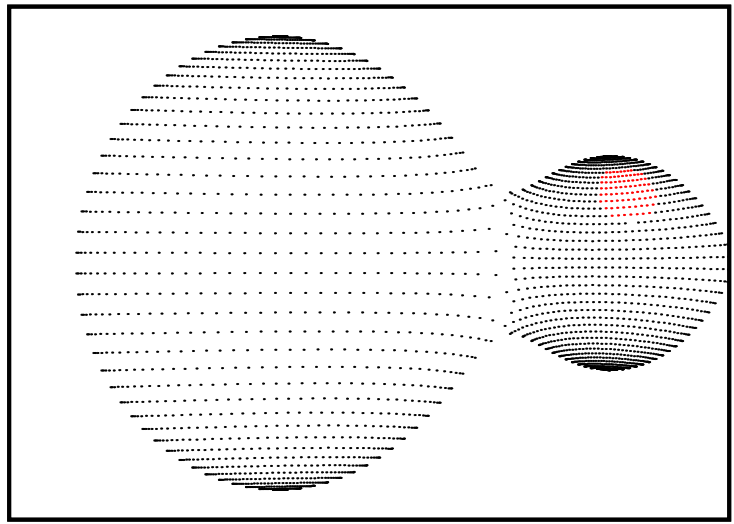}
\caption{The geometric configuration of V728 Her at $\phi$=0.25.}
\label{figgeo}
\end{figure}

In order to represent the magnitude difference between two consecutive maxima we decided to add a cool spotted region on the surface of the one of the stars.  Since the  magnetic activity of the secondary component is expected to be more efficient (see  Sec.~2) the cool spot is assumed to be on the surface of the secondary companion. The best spot parameters were reached at the smallest standard deviation of the simultaneous solution. Furthermore, we applied an alternative solution by setting the third light parameter $l_3$ as adjustable following to our results from the period study. The results of the light curve analyses are listed in Table~\ref{tablc} in which SI and SII refer to the analyses with and without the assumption of tertiary component, respectively. Fig.~\ref{figlc} shows the theoretical light and radial velocity curves calculated in SI. The geometric configuration of the system is also presented in Fig.~\ref{figgeo}.
 
When it comes to compare our results to previous studies, our $q$ value of 0.156, is slightly different from the convective model result of \cite{nel95}, 0.178. The temperature differences given in previous studies are 154~K in \cite{nel95} and 98~K in \cite{sam90} while we calculated the difference between hotter (less massive) and cooler (more massive) components as 143~K. Our orbital inclination value, 69$^\circ$.3, is close to 68$^\circ$.14 of \cite{nel95}, however, it differs considerably from the inclination (64$^\circ$.75) derived by \cite{sam90}. We calculated the fillout factor, 81\%, which is larger than the value of \cite{nel95}, 71\%. The difference between this factor mainly arises from the different mass ratios derived in these two solutions.

\section{Discussion and Conclusion}

\begin{figure}
\centering
\includegraphics[scale=1.2, angle=0]{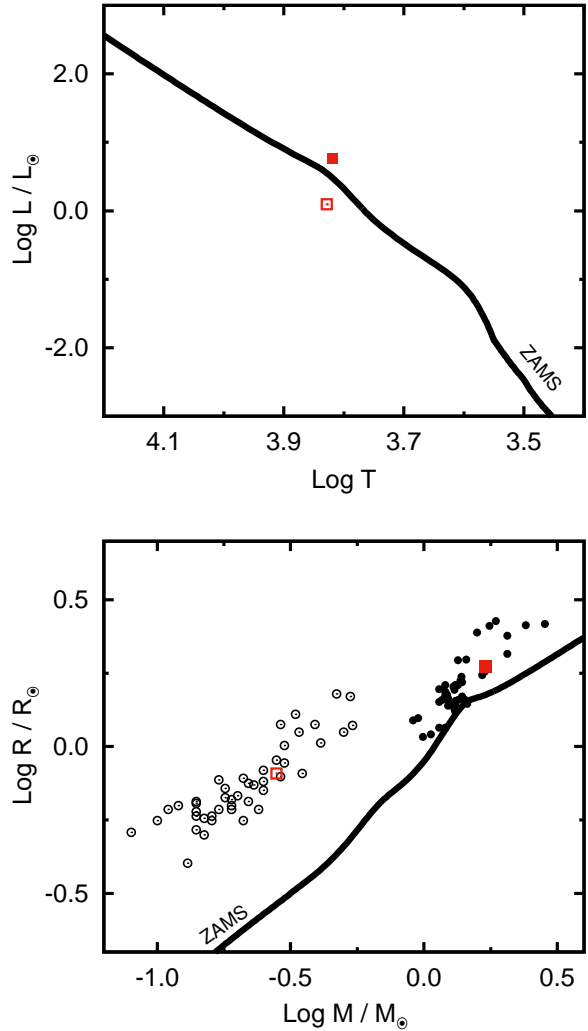}
\caption{The components of the system in the Hertzsprung--Russell diagram (top panel) and mass--radius plane (bottom panel) together with other low--mass deep--contact systems. The filled and open circles are stand for the primary and secondary components, respectively. The components of \astrobj{V728~Her} are symbolized by squares. ZAMS data are taken from \cite{pol95}. See text for further details.}
\label{fighr}
\end{figure}

The new $BVR$ light curves of the binary system \astrobj{V728~Her} is combined with the radial velocity data of \citep{nel95} and are solved simultaneously. The derived physical parameters of the system are given in Table \ref{tababs}. These parameters are also used to locate the components in the Hertzsprung--Russell diagram (Fig.~\ref{fighr}). The primary component of the system is situated above the Zero Age Main Sequence (ZAMS) whereas the secondary one located below it.

The variation in orbital period of the binary was investigated by analyzing the collected times of minimum light. The results show that the parabolic variation is the consequence of conservative mass transfer from the less massive component to the other one with a rate of $2.51\times10^{-8}$ M$_{\odot}$yr$^{-1}$. The additional sine--like variation superposed on the parabola was also examined and the results were interpreted in terms of two probable physical phenomena. First, the light--time effect because of the existence of a tertiary component: the system might have an external component having a mass about 0.4M$_{\odot}$. Second, the Applegate mechanism which causes a modulation in orbital period related to the magnetic activity of one of the components. Our results for the second model imply that there is a probability for the secondary component to stimulate the Applegate mechanism.

We derived the degree of contact of the system, $f$, using the equation $f=\frac{\Omega_{i}-\Omega}{\Omega_{i}-\Omega_{o}}$ \citep{luc79}, where $\Omega_{i}$ is the inner and $\Omega_{o}$ is the outer Lagrangian potentials. The result, $f$=81\%, indicates that the system is a deep contact binary according to the criteria ($f\geqslant$~50\%) given by \cite{qia05a}. In addition, the morphology of the light curve shows that the system is a W--type W~UMa binary according to the classification of \cite{bin70} who categorized the W~UMa systems into two sub-classes: A--type systems in which the more massive star eclipsed during the deeper minimum and W--type systems where the less massive component is hotter than the more massive one. However, the radial velocity curve of \cite{nel95} indicates that the massive star is occulted during the primary minimum, thus the situation is contrary to classifying the binary as W--type W~UMa system. Furthermore, \cite{yil13} noticed that the mass--luminosity relation and the angular momentum properties of the system require to put the binary in A--type subclass. The authors also stated that the initial mass of the secondary component is higher than that of A--type secondaries.

\cite{zhu11} remarked that deep, low mass ratio binary systems with increasing orbital period are in the late--type tidally locked stages of their evolution. These stars may evolve from their present situation to rapidly rotating single stars. Thus, the low mass ratio contact binaries with short period are potential samples for explaining the merging phenomena in binary star systems \citep{arb09,jia10}. \cite{yan15} investigated the relations among the physical parameters of deep, low mass ratio contact systems. They concluded that these systems may evolve to rapidly rotating single stars following the inner and outer Roche Lobe shrinkage phase. \cite{ras95} and \cite{ras94,ras95b} mentioned that the contact systems having relatively low mass ratios ($<$0.45) and deep convective envelopes are entering their final phase of merger. Following \cite{ras95} and \cite{ras94,ras95b}, \cite{nel95} noted that \astrobj{V728~Her} would be at the final stage of merger if it shows a period change. As we proved the orbital period change, found the large fill-out factor ($f$=81\%) and applied a simultaneous solution fitting the convective model assumption the system could be close to reaching a merger. However, the gyration radius ($k^2$=0.06, smaller than 0.16 of \cite{nel95}) of the primary component falls into stable region on the diagram given by \cite{ras95}. 

In order to compare \astrobj{V728~Her} to deep, low mass ratio contact binaries ($q\leqslant$~0.25,~$f\geqslant$~50\%) we plotted the system on the mass--radius plane (Fig.~\ref{fighr}) with the other deep, low mass ratio contact systems whose physical parameters were given by \cite{yan15}. It can be clearly seen from the figure that the components of our target are in a good agreement with the other low mass ratio contact binary systems.

\section*{Acknowledgements}
The authors acknowledge their thanks to Dr. Petr Zasche for his valuable comments on running the {\tt LITE} code.

\newpage

\begin{table}
\centering
\caption{Calculated times of minimum light.}
\label{tabmin}
\begin{tabular}{ll}
\hline
\multicolumn{1}{c}{HJD} & Type      \\
\hline
2456485.47292(1)           & \multicolumn{1}{c}{I} \\
2456489.48011(1)           & \multicolumn{1}{c}{II}\\
\hline
\end{tabular}
\end{table}

\begin{table}
\centering
\caption{Final results of the orbital period analysis. $P^\prime$ denotes the period of tertiary component. $A$ and $f(m)$ refer to the semi--amplitude of the sinusoidal variation and the mass function. Formal error estimates are given in parenthesis.}
\label{tabth}
\begin{tabular}{ll}
\hline
Parameter                                & Value   \\
\hline
$T_0$~(HJD)                              & 2446949.845(2)  \\
$P_0$~(d)                                & 0.4712889(1)  \\
$Q$					& 1.2438(8)10$^{-10}$ \\
$P^\prime$~(yr) 			    & 75.9(9) \\
$a _{12}\sin i^\prime$~(AU)                 & 3.9(1.9)    \\
$A$~(d)    & 0.02(1) \\
$f(M)$~(M$_{\odot}$)                        & 0.01(4)   \\
\hline
\end{tabular}
\end{table}

\begin{table}
\centering
\scriptsize
\caption{Result of the light curve analysis. SI refers to the solution which contains contribution of third light while SII denotes the analysis without assumption of tertiary component. Spot parameters are co--latitude $\beta$, longitude $\lambda$, fractional radius $r$ and the temperature factor $t$. Formal error estimates are given in parenthesis.}
\label{tablc}
\begin{tabular}{lcc}
\hline
Parameter                                   & \multicolumn{2}{c}{Value}   \\
 & SI & SII \\
\hline
$i$ ${({^\circ})}$                          & 69.3(1)  &69.4(1)\\
$q$                                         & 0.158(1) &0.156(1) \\
$V_{0}$~(km/s)				    & 43.4(7)  &43.4(7)	\\
$a$ (R$_\odot$)				    & 3.22(2)  & 3.22(3)\\
$\Omega _{1}=\Omega _{2} $                  & 2.024(1) & 2.028(6)   \\
$T_1$~(K)                               & \multicolumn{2}{c}{6600}   \\
$T_2$~(K)                               & 6743(27) & 6640(21)\\
Fractional radius of primary                & 0.5811(2) & 0.5788(3)\\
Fractional radius of secondary              & 0.2935(58)& 0.2832(17) \\
Luminosity ratio:$\frac{L_1}{L_1 +L_2 +l_3}$ &   \\
$B$                                         & 0.752(5)  &0.815(2)      \\
$V$                                         & 0.783(5)  &0.817(2)      \\
$R$                                         & 0.778(5)  &0.818(2)       \\
Spot parameters: & \\
$\beta$ ($^{\circ}$)    & \multicolumn{2}{c}{50} \\
$\lambda$ ($^{\circ}$)  & \multicolumn{2}{c}{110} \\
$r$              & \multicolumn{2}{c}{20} \\
$t$              & \multicolumn{2}{c}{0.9} \\
Luminosity ratio:$\frac{l_3}{L_1 +L_2+l_3}$ &   \\
$B$                                         & 0.056(4)   &     \\
$V$                                         & 0.028(4)   &     \\
$R$                                         & 0.034(4)   &      \\
\hline
\end{tabular}
\end{table}

\begin{table}
\centering
\caption{Absolute parameters of the system derived from the light and radial velocity curve solution results. The effective temperature of the sun is set to 5780~K.}
\label{tababs}
\begin{tabular}{lrr}
\hline
Parameter                &\multicolumn{2}{c}{Value}    \\
& \multicolumn{1}{c}{Pri.}&\multicolumn{1}{c}{Sec.} \\
\hline
M (M$_{\odot}$)          & 1.8(1)         & 0.28(8) \\
R (R$_{\odot}$)          & 1.87(4)          & 0.82(6)\\
L (L$_{\odot}$)		 & 5.9(3)	     & 1.24(2) \\
T (K)			 & 6600		     & 6743(27) \\
M$_{bol}$ ($^m$)         & 2.8(3)	     & 4.5(9) \\
$a$ (R$_{\odot}$)        & 0.44(5)           & 2.78(5) \\
\hline
\end{tabular}
\end{table}

\newpage

\newpage

\newpage

\newpage

\end{document}